\documentclass[9pt,twocolumn,twoside]{opticajnl}
\journal{opticajournal} 

\setboolean{shortarticle}{true}

\usepackage{subcaption}
\usepackage{graphicx}

\usepackage{lineno}
\linenumbers 

\usepackage{amsmath}
\newcommand{\murm}{%
    \mathchoice
        {\hbox{\normalsize\textmu}}
        {\hbox{\normalsize\textmu}}
        {\hbox{\scriptsize\textmu}}
        {\hbox{\tiny\textmu}}%
}

\usepackage[font=small,skip=0pt]{caption}

\title{Simulation of diffraction and scattering using the Wigner Distribution Function}

\author[1]{Emilie Pietersoone}
\author[2]{Jean Michel Létang}
\author[2]{Simon Rit}
\author[1]{Max Langer}

\affil[1]{Univ. Grenoble Alpes, CNRS, UMR 5525, VetAgro Sup, Grenoble INP, TIMC, F-38000 Grenoble, France}
\affil[2]{Univ. Lyon, INSA-Lyon, Université Claude Bernard Lyon 1, UJM-Saint Étienne, CNRS, Inserm, CREATIS UMR 5220, U1294, F-69373 Lyon, France}

\begin{abstract}X-ray phase-contrast imaging enhances soft tissue visualization by leveraging the phase shift of X-rays passing through materials. It permits to minimize radiation exposure due to high contrast, as well as high resolution imaging limited by the wavelength of the X-rays. Phase retrieval extracts the phase shift computationally, but simulated images fail to recreate low-frequency noise observed in experimental images. To this end, we propose a new method to simulate phase contrast images using the Wigner Distribution Function. This permits the simulation of wave and particle effects simultaneously and simulates images photon by photon. Here, we give a first demonstration of the method by simulating the Gaussian double-slit experiment. It has the potential for realistic simulation of low-dose imaging. 
\end{abstract}

\setboolean{displaycopyright}{false} 

\begin{document}

\nolinenumbers

\maketitle

Conventional X-ray imaging relies on the attenuation of X-rays for contrast. X-ray phase contrast, on the other hand, achieves contrast from the phase shift of the beam by the sample. 
There are several ways to achieve phase contrast, for example by a grating interferometer or a spatial modulation mask. The simplest way to achieve phase contrast, however, is to let the beam propagate in free space after interaction with the sample, which is known as in-line phase contrast or propagation-based imaging. 

The main interest in phase contrast imaging is the increased contrast in soft materials, or between materials with similar electron densities. In-line phase contrast also arises in high-resolution imaging using X-ray optics \cite{mokso2007nanoscale, langer2012x}. The increased contrast allows both to improve visualisation of these structures and to reduce the imaging dose, important both in medical applications and in high-resolution imaging. 

The main inconvenience is that the phase shift is not directly accessible in the detector plane. The phase contrast contains entangled information from both the attenuation and the phase shift. The phase (and potentially the attenuation) must instead be reconstructed from the phase contrast images, a process known as phase retrieval. 
\begin{figure}[!t]
    \centering
    \includegraphics[width=\columnwidth]{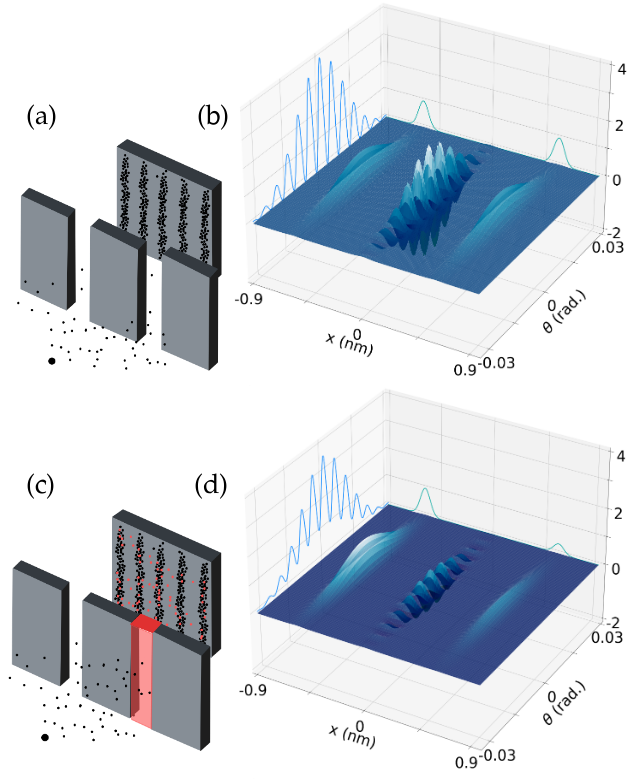}
    \caption{Numerical experiments and their corresponding Wigner distribution function (WDF). (a) The double-slit experiment is modelled by two Gaussians. (b) Analytic WDF and its marginals, corresponding to the intensity in the spatial and angular domains, respectively. (c) Modified double-slit experiment with a scattering element in one of the slits. (d) Corresponding WDF with the amplitude in the slit with the scatterer set to 0.35.}
    \label{fig:double-slit_simulation}
\end{figure}
One problem that persists in phase retrieval is the occurrence of noise in the low spatial frequency range, giving rise to cloud-like artefacts in the reconstructed images. Several algorithms have been proposed to address this problem on the reconstruction side \cite{paganin2002, langer2010, langer2014, langer2012OL} but a clear understanding of the origin of these artefacts has not been established. While it is well known that phase contrast is less sensitive to low spatial frequencies in the phase shift, thus making phase retrieval sensitive to noise in this range, this does not fully explain the qualitative appearance of the noise in the reconstructed images. Indeed, current simulation techniques are unable to reproduce these kinds of artefacts. Instead, reconstructing the phase from simulated images yields a noise with the appearance of low-pass filtered noise \cite{langer2008quantitative, paganin2004quantitative, instruments8010008}. 

To address this problem, we propose a new way of simulating phase contrast images. We hypothesise that the observed low-frequency artefact stems from incoherent scattering in the sample. The aim is therefore to combine phase contrast, originating from diffraction - a wave effect, and incoherent scattering - a particle effect, into one simulation. We propose to do this by using the Wigner Distribution Function (WDF) of the wave exiting the sample. As an initial demonstration, we simulate a Gaussian double-slit experiment, as well as a modified double-slit experiment that includes scattering. 

The popularity of phase contrast imaging can mainly be explained by the higher sensitivity compared to standard attenuation-based X-ray imaging. The complex refraction for hard X-rays consists of two components, one related to phase shift ($\delta$) and one to attenuation ($\beta$) \cite{goodman2005introduction, paganin2006coherent}: 
\begin{equation}
\label{eq:refractive_index}
    n(x,y)=1-\delta(x,y)-i\beta(x,y)
\end{equation}
The higher sensitivity is due to that, depending on the X-ray energy and the imaged material, $\delta$ can be several orders of magnitude greater than $\beta$ depending on the wave length $\lambda$ \cite{langer2008quantitative}. 
The phase $\phi$ and amplitude $B$ of the wave are usually considered straight-line projections through the refractive index 
\begin{equation}
\varphi(x)=-\frac{2 \pi}{\lambda}\int \delta (x,y) \mathrm{d}y, \, B(x)=\frac{2 \pi}{\lambda}\int \beta (x,y) \mathrm{d}y,
\label{eq:phase_shift_projection}
\end{equation}
%
%
so that the exit wave 
is $\Psi(x) = \exp [-B(x)] \exp [i\varphi(x)]$.
The usual model for phase contrast when the propagation distance $D$ is relatively short is Fresnel diffraction. In the Fresnel model, the contrast $I_D(x)$ on the detector can be written as
\begin{equation}
I_D(x)\approx \left | \int \Psi(x) \exp\left(i\frac{\pi}{\lambda D} |x|^2 \right ) \right |^2.
\label{eq:Fresnel}
\end{equation}
When the propagation distance is longer than a certain distance (Fraunhofer distance), the contrast is modelled by Fraunhofer diffraction, which is essentially the amplitude of the Fourier transform of the exit wave:
\begin{equation}
I_D(x)\approx\left | \mathcal{F} \{ \Psi \}\left( \frac{x}{\lambda D} \right) \right |^2.
    \label{eq:Fraunhofer}
\end{equation}

The most straight-forward way to simulate phase contrast is to directly implement Eqs. \ref{eq:Fresnel} and \ref{eq:Fraunhofer} numerically. Noise is added \textit{a posteriori} to the ideal diffraction images. While this approach is simple to implement, it cannot account for effects such as reflection, refraction and incoherent scattering; and when reconstructing the phase from such images, the result does not look like the artefacts observed in experimental images (e.g., \cite{langer2010, suuronen2022, weber2016}). The noise looks like low-pass filtered noise instead of the slowly varying, cloud-like artefacts seen in experimental images \cite{paganin2004quantitative, langer2008quantitative, quenot_evaluating_simulators}.

Several approaches combining phase contrast and incoherent scattering have recently been proposed. These are reviewed in \cite{chen, bravin, berujon, quenot_review, quenot_evaluating_simulators, instruments8010008}. They generally depend on a wave-optics propagation combined to a Monte Carlo simulation for the scattering, and are either two-part simulations or combined methods. Bartl et al. \cite{bartl2009simulation} introduced a method that combines a wave simulation with a particle simulation, encompassing all relevant physical aspects, including the behaviours of X-ray photons as particles and waves, realistic noise, and a comprehensive description of the X-ray source and detector. Peter et al. \cite{peter} introduced a simulation framework for phase-sensitive X-ray imaging that accounts for both the particle and wave characteristics of X-rays. This split approach involves a three-step simulation, rendering the framework flexible and suitable for various methods of phase-sensitive X-ray imaging. Tessarini et al. \cite{tessarini2022semi} introduced a simulation framework for phase-sensitive X-ray imaging, which relies on semi-classical properties of particles, enabling the integration of both particle and wave-like effects. Langer et al. \cite{langer2020towards} incorporated a Monte Carlo (MC) process to simulate the refraction and total reflection of X-rays. This method ensures a detailed and comprehensive simulation of X-ray Phase Contrast Imaging (XPCI) within GATE. The simulation covers absorption, scattering, and refraction, seamlessly integrating wave optics into the GATE framework. The implementation of refraction utilizes a boundary process within GATE, providing adaptability to various imaging methods through adjustments to specific source and detector parameters.


Here, we seek a way to account for both coherent and incoherent effects in the same simulation. To do this, we propose to use the WDF to calculate the interference in the object plane rather than in the detector plane, as is done in the classical methods. This allows to convert the exit wave into position and momentum and thus a particle-type process that can be combined with incoherent scattering. 

\begin{figure}[!t]
\centering
\begin{subfigure}{.49\columnwidth}
  \centering
    \includegraphics[width=\textwidth]{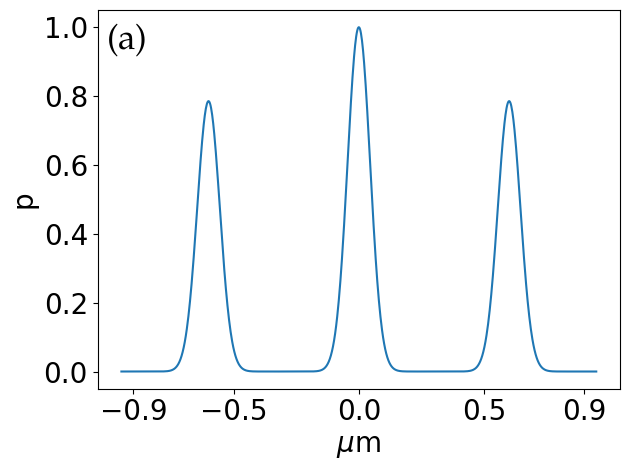}
\end{subfigure}
\begin{subfigure}{.49\columnwidth}
    \centering    
    \includegraphics[width=\textwidth]{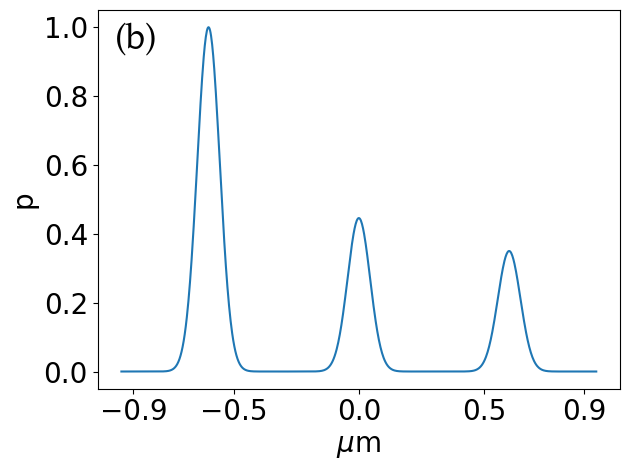}
\end{subfigure}
\caption{The spatial probability with interference ($S\psi$, Eq. \ref{eq:spi}) gives the probability of finding a photon in a certain point in the slit plane taking into account the interference term. (a) Equal amplitudes. (b) Right slit amplitude set to $0.35$ as in (Fig. \ref{fig:double-slit_simulation}b).}
\label{fig:SPIs}
\end{figure}

The WDF has previously been applied to optics to different ends \cite{mout, alonso}. For our purposes, the WDF can be written as
\begin{equation}
    \label{eq:wigner_optics}
    W_\Psi (x, \theta ) = \int \Psi \left( x+x' \right) \Psi^* \left( x-x' \right) \mathrm{exp} \left(- i2 \pi x' \theta \right) \mathrm{d}x'.
\end{equation}

The WDF is a quasi-probability density function, meaning it represents the joint probability distribution of positions $x$ and angles $\theta$, but it can take negative values, representing destructive interference in the diffracted wave. This is clearly seen in the WDF of two Gaussians, (Fig. \ref{fig:double-slit_simulation}), where the WDF has an oscillatory interference component between the two Gaussians. The marginal along $x$ corresponds to the intensity of the Fourier transform of $\Psi(x)$
\begin{equation} \label{eq:x_marginal}
    |\mathcal{F}\{\Psi\}(\theta)|^2= \int W_\Psi(x,\theta) \mathrm{d}x
\end{equation}
which is the intensity past the Fraunhofer distance. 
The marginal along $\theta$ gives the intensity at the exit of the slits
\begin{equation} \label{eq:theta_marginal}
    I_0(x)=|\Psi(x)|^2= \int W_\Psi(x,\theta) \mathrm{d}\theta.
\end{equation}


Inspired by the negative particle formulation of quantum physics \cite{sellier2015}, the idea is then to generate particle trajectories from the WDF by generating positions and corresponding angles probabilistically. To do this, we first need the probability of a photon in space taking into account the interference term. Since, in the projection over $\theta $, the positive values will cancel out the negative \cite{bartlett1945}, we instead project the modulus of the WDF
\begin{equation} \label{eq:spi}
    S_\Psi(x)=\int |W_\Psi(x,\theta)| \mathrm{d}\theta.
\end{equation}

To account for scattering in a simple way, we model it by introducing a block of material of a certain thickness and scattering coefficient, which are set to yield a certain scattering probability $p_s$ within the support of the block, and is for this demonstration considered thin (one scattering interaction at most, and all scattering occurs in the plane of the  slits). If a particle is scattered, its scattering angle is calculated by probability sampling from the differential cross-section of the chosen material. 

A position $x_n$ is drawn at random with uniform probability. A random value $ \epsilon_s \in [ 0, 1 ] $ is drawn, and if $ \epsilon_s < p_s $ the photon is considered to be scattered. The scattering angle is generated from the DCS, the particle is then ray-traced in a straight line to the detector and one count is registered in the corresponding detector element. If the particle was not scattered, another random number $\epsilon_d$ is generated. If $ \epsilon_d < S(x_n) $ we consider the photon in $x_n$ was diffracted, otherwise it is considered absorbed or otherwise annihilated. We then draw a diffraction angle $\theta_n$ by sampling from the probability distribution $P_\theta (x_n)=|W(x_n,\theta)|/\sum |W(x_n,\theta)|$ and the photon is ray-traced in a straight line to the detector. The sign of the particle is retrieved from $W_\Psi(x_n,\theta_n)$. If the sign is negative, a negative potential is incremented to account for future destructive interference. If the sign is positive, it either decrements the negative potential in that pixel or, if the potential was zero, registers a count on the detector. 
As a first demonstration of the proposed method, a Gaussian double-slit (GDS) experiment is simulated. This is chosen since the GDS has an analytical solution for the WDF, which for now enables us to bypass the most challenging numerical difficulties in this initial demonstration. Using Gaussian functions reduces the oscillatory nature of the interference term. This is done mainly for clarity; the demonstration works also for the standard double-slit. To include scattering, a scattering element is placed in one of the slits (Fig. \ref{fig:double-slit_simulation}c), with uniform scattering probability. 
\begin{figure}[!t]
\centering
\begin{subfigure}{.5\columnwidth}
  \centering
    \includegraphics[width=\textwidth]{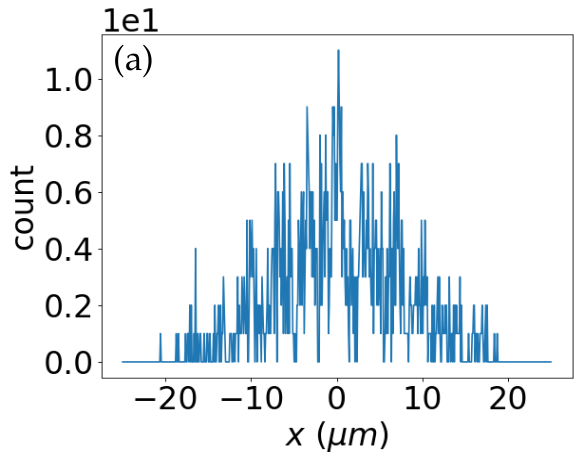}
\end{subfigure}
\begin{subfigure}{.49\columnwidth}
  \centering
    \includegraphics[width=\textwidth]{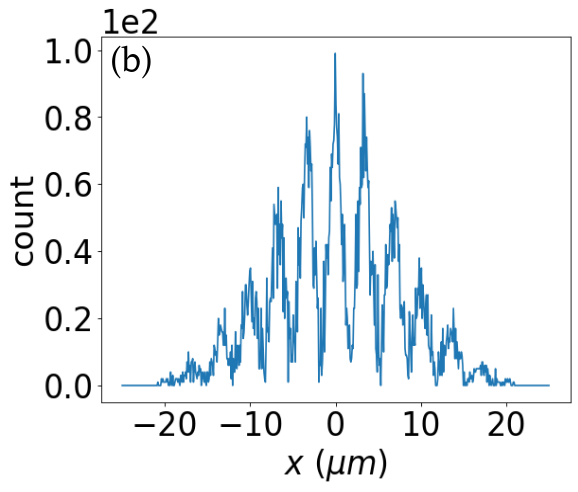}
\end{subfigure}
\begin{subfigure}{.49\columnwidth}
  \centering
    \includegraphics[width=\textwidth]{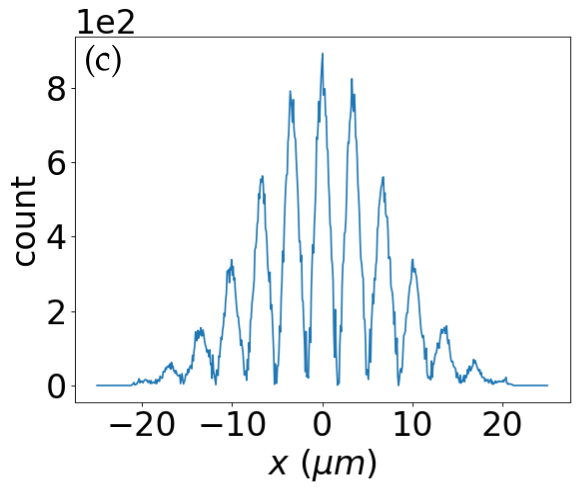}
\end{subfigure}
\begin{subfigure}{.49\columnwidth}
  \centering
    \includegraphics[width=\textwidth]{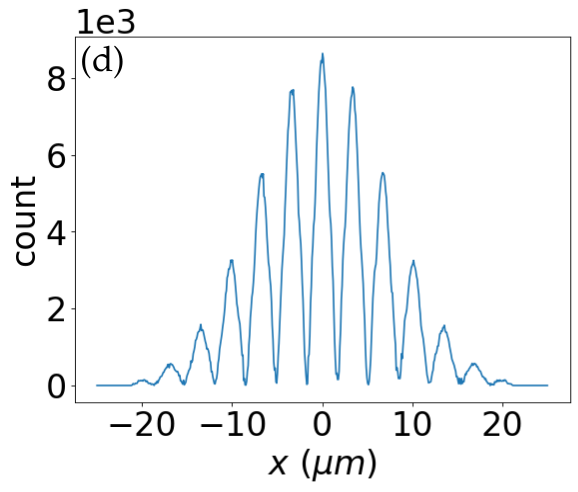}
\end{subfigure}
\caption{Simulations of the GDS for different photon counts at $D=650 \, \mu m$: (a) $10^3$ (b) $10^4$ (c) $10^5$  (d) $10^6$ photons.} 
\label{fig:number_of_photons}
\end{figure}
The exit wave from the GDS can be written as \cite{cuypers2011validity, cervenka2021deterministic}
\begin{equation}
    \Psi_{GDS}(x)= B_1 e^{ - \frac{(x-a)^2}{A^2} } + B_2 e^{ - \frac{(x+a)^2}{A^2} }, 
    \label{eq:gaussian_young_mod}
\end{equation}
where we have added factors $B_1$ and $B_2$ to account for the attenuation caused by the scattering element, $a$ is the position of the Gaussians (symmetrically around $0$), and $A$ their standard deviation.
%
%
The WDF of $\Psi_{GDS}(x)$ can then be written as
\begin{equation} 
    \begin{split}
     W_{\Psi_{GDS}} (x, \theta ) &= B_1 e^{ - \frac{(x-a)^2}{A^2} } e^{ - \left( \frac{\theta}{\lambda} \right)^2 A^2 } + B_2 e^{ - \frac{(x+a)^2}{A^2} } e^{ - \left( \frac{\theta}{\lambda} \right)^2 A^2 } \\
        &+ B_1 B_2 2 e^{ - \frac{x^2}{A^2} } e^{ - \left( \frac{\theta}{\lambda} \right)^2 A^2 } \cos \left( \frac{2a\theta}{\lambda} \right),
    \end{split}
    \label{eq:wdf_young_gauss_mod}
\end{equation}
where we have omitted the scaling factor that would cause the function to be a probability density (this aspect is actually never used here). This function is shown in Fig. \ref{fig:double-slit_simulation} in two configurations along with its marginals. 
In this experiment, the slit positions were set to $a=0.6$~{\textmu}m, the standard deviation to $A=60$~nm, and the wavelength to $\lambda=1$~nm. The detector had the size $250$~{\textmu}m and had 601 pixels. 

First, the simulation was run without a scattering element. $S_\Psi(x)$ for this configuration is shown in Fig. \ref{fig:SPIs}(a). Results for increasing number of photons are shown in Fig. \ref{fig:number_of_photons}. The building up of the diffraction pattern photon by photon is shown in the supplementary video S1.

\begin{figure}[!t]
    \small
    \centering
    \includegraphics[width=\columnwidth]{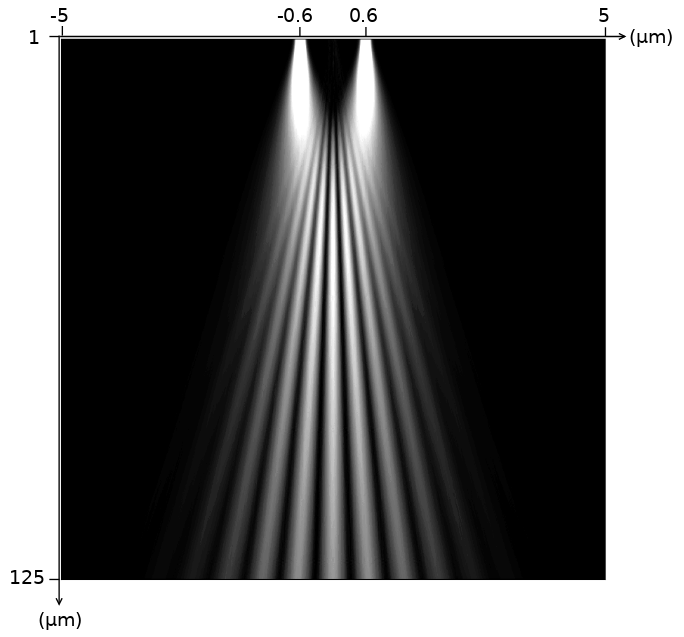}
    \caption{Evolution of the intensity with 601 propagation distances over $D \in [1 \, \murm\mathrm{m}, 125 \, \murm\mathrm{m}]$ with a detector width of $[-5 \, \murm\mathrm{m}, 5 \, \murm\mathrm{m}]$. As expected, close to the slits no interference is visible. With sufficient propagation, interference fringes become visible and evolve with the propagation distance (near-field). Beyond a certain distance, the pattern stops evolving qualitatively and only spreads more in space.}
    \label{fig:talbot}
\end{figure}

\begin{figure}[!t]
\centering
\begin{subfigure}{.49\columnwidth}
  \centering
    \includegraphics[width=\textwidth]{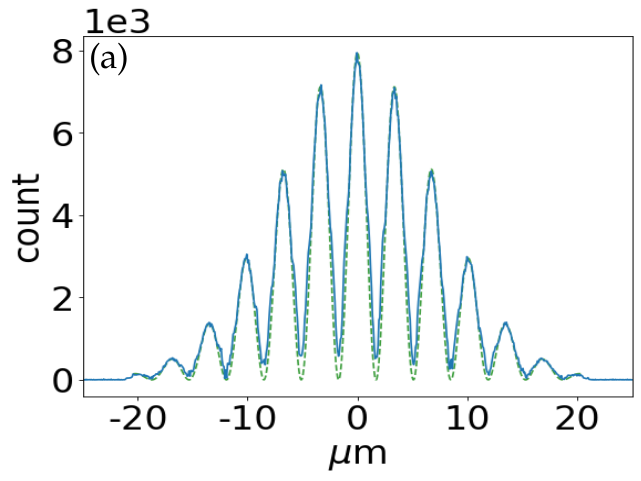}
\end{subfigure}
\begin{subfigure}{.49\columnwidth}
  \centering
    \includegraphics[width=\textwidth]{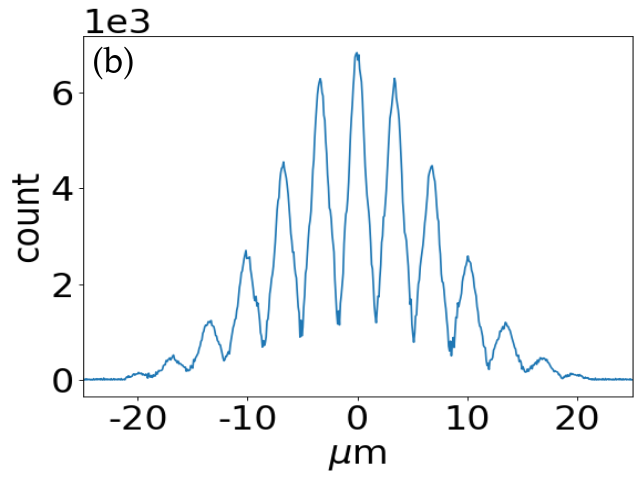}
\end{subfigure}
\begin{subfigure}{.49\columnwidth}
  \centering
    \includegraphics[width=\textwidth]{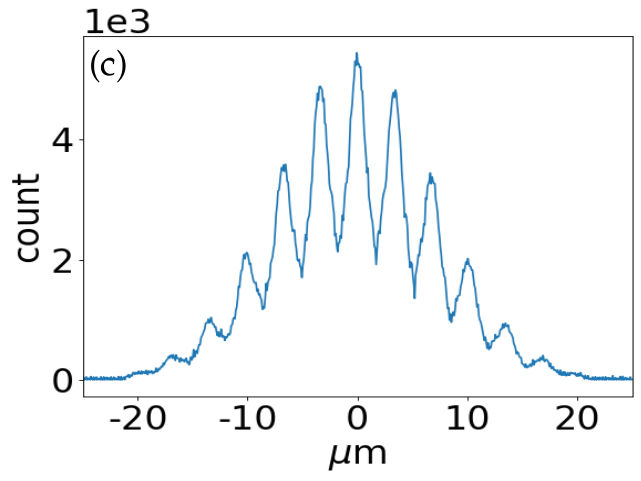}
\end{subfigure}
\begin{subfigure}{.49\columnwidth}
  \centering
    \includegraphics[width=\textwidth]{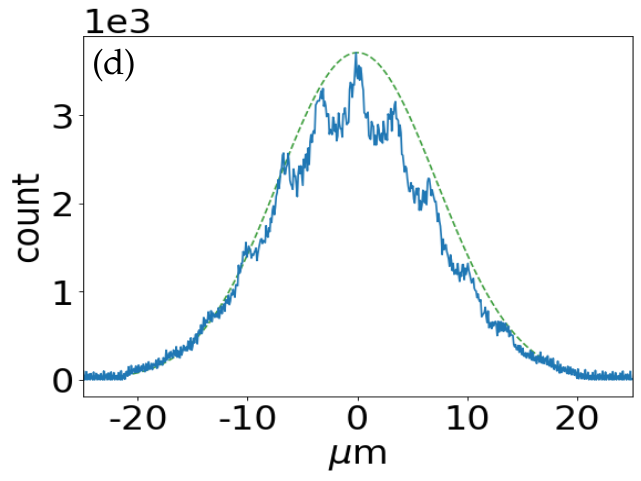}
\end{subfigure}
\caption{Simulations of the modified GDS framework for $10^6$ photons at $D=650$~{\textmu}m for different scattering probabilities: (a) 25\% (theoretical double-slit pattern shown in dashed green) (b) 50\% (c) 75\% (d) 90\%, (theoretical single-slit pattern shown in dashed green).}
\label{fig:scattering_tests}
\end{figure}

Further, this simulation was run over a range of propagation distances to demonstrate the capability to simulate the diffraction pattern regardless of the propagation distance. In this simulation, the detector size was set to 10~{\textmu}m. The evolution of the intensity with distance is shown in Fig. \ref{fig:talbot}. As expected, the contrast evolves in the near-field, then at a certain distance the far-field pattern emerges and remains stationary except for a scaling.
To test the combination of coherent and incoherent effects, a scattering element was introduced in one of the slits. To account for the attenuation due to the scattering, the amplitude $B_2$ in Eqs. \ref{eq:gaussian_young_mod} and \ref{eq:wdf_young_gauss_mod} was adjusted accordingly. The effect on the WDF and its marginals is shown in Fig. \ref{fig:double-slit_simulation}b, and the effect on $S_\Psi(x)$ is shown in Fig. \ref{fig:SPIs}. 
To calculate scattering coefficients and scattering angles, we used the Rayleigh cross-section for Si ($Z=14$, $\rho=2.393$~g.cm$^{-3}$). The thickness of the material was varied to yield different scattering probabilities.
The differential cross-section (DCS) is used as the scattering angular distribution and is sampled in $\theta \in [-\frac{\pi}{2}, \frac{\pi}{2}]$ using 9001 sampling points. Simulation with different scattering probabilities showed that when increasing the attenuation in one slit due to scattering, the diffraction pattern approaches the single-slit pattern, as expected (Fig. \ref{fig:scattering_tests}).


We presented a new approach to simulate X-ray phase contrast images, and by extension coherent imaging systems in general by the use of the WDF and the signed particle formulation of quantum physics. This approach offers several interesting aspects, namely that it can combine the simulation of coherent and incoherent effects, it is independent of the propagation distance, and it builds up the images photon by photon. While the original motivation for this work was to combine coherent and incoherent effects in one simulation, the last aspect is especially interesting in view of recent developments of photon counting detectors. Recently, the double-slit experiment was demonstrated with X-rays using a photon counting detector \cite{gureyev2024young}, giving further weight to this argument.    


The main limitation with this initial study is that the WDF is known analytically. Calculating the WDF numerically is notoriously difficult. This is due to its large size, effectively doubling the dimensionality of the signal, its generally highly oscillatory nature \cite{alonso}, and the difficulties to correctly discretize it \cite{bjork2008}. This choice was made to permit this demonstration while temporarily avoiding these difficulties. The method has several redeeming features however, making its numerical and practical implementation seem feasible. The WDF never needs to be calculated in its entirety. It can be calculated separately point by point in space, first to generate $S_\Psi(x)$ once at the beginning of the simulation, then to generate $P_\theta (x_n)$ for each particle. This can be done in a parallel manner for each particle, for example as a part of a particle transport code. The oscillatory nature of the WDF could possibly be controlled by imposing some kind of regularity on the exit wave, or by attempting to localize interference and sample accordingly. 

If these difficulties can be overcome, the method has the potential to yield realistic simulations in low-dose imaging scenarios. The simultaneous simulation of scattering and phase contrast would allow to address more precisely the low-frequency noise problem in phase retrieval, for example through providing data for machine learning-based algorithms.

To conclude, we have presented a new method to simulate phase-contrast images taking into account both coherent and incoherent effects. While the first numerical results are encouraging, future work will focus on the appropriate numerical implementation of the WDF in this context to permit a practical implementation, as well as experimental validation of the method.





\begin{backmatter}

\bmsection{Funding} This research was funded by the French Agence Nationale de la Recherche, grants MIAI@Grenoble Alpes (ANR-19-P3IA-0003), Labex CAMI (ANR-11-LABX-0004) and (Labex PRIMES ANR-11-LABX-0063). 

\bmsection{Acknowledgments} We thank Emmanuel Brun, STROBE, Université Grenoble Alpes, for fruitful discussion and suggestions. 





\bmsection{Disclosures} The authors declare no conflicts of interest.

\bmsection{Supplemental document}
See Supplement 1 for supporting content. 

\end{backmatter}




\bibliography{Optica-journal-template}

\bibliographyfullrefs{Optica-journal-template}


\ifthenelse{\equal{\journalref}{aop}}{%
\section*{Author Biographies}
\begingroup
\setlength\intextsep{0pt}
\begin{minipage}[t][6.3cm][t]{1.0\textwidth} 
  \begin{wrapfigure}{L}{0.25\textwidth}
    \includegraphics[width=0.25\textwidth]{john_smith.eps}
  \end{wrapfigure}
  \noindent
  {\bfseries John Smith} received his BSc (Mathematics) in 2000 from The University of Maryland. His research interests include lasers and optics.
\end{minipage}
\begin{minipage}{1.0\textwidth}
  \begin{wrapfigure}{L}{0.25\textwidth}
    \includegraphics[width=0.25\textwidth]{alice_smith.eps}
  \end{wrapfigure}
  \noindent
  {\bfseries Alice Smith} also received her BSc (Mathematics) in 2000 from The University of Maryland. Her research interests also include lasers and optics.
\end{minipage}
\endgroup
}{}

\end{document}